\documentclass[a4paper]{jpconf}
\usepackage{graphicx}
\begin{document}
\title{Spin-Orbital Entanglement Emerging from Frustration in the Kugel-Khomskii Model }

\author{ W Brzezicki $^1$ and A M Ole\'{s} $^{1,2}$ }

\address{$^1$ Marian Smoluchowski Institute of Physics, Jagellonian
              University, \\ Reymonta 4, PL-30059 Krak\'ow, Poland \\
         $^2$ Max-Planck-Institut FKF,
              Heisenbergstrasse 1, D-70569 Stuttgart, Germany}

\ead{w.brzezicki@uj.edu.pl}

\begin{abstract}
We investigate the zero-temperature phase diagrams of the 
bilayer square-lattice Kugel-Khomskii ($d^9$) model involving entangled 
and singlet phases using mean-field cluster approach. This diagram includes 
interlayer singlet phase observed in fluoride K$_3$Cu$_2$F$_7$ and
exotic entangled spin-orbital phases. 
For a monolayer case, realized in K$_2$CuF$_4$, we perform similar
calculations in finite temperature and show that 
the alternating-orbital ferromagnet decays first to an
entangled uniform ferromagnet and then to a paramagnetic
phase. \\
{\hskip .3cm \it Published in: J. Phys.: Conf. Series} {\bf 391}, 012085 (2012).
\end{abstract}

It has been shown that quantum fluctuations are enhanced near the
orbital degeneracy and could suppress long-range order in the
Kugel-Khomskii (KK) model \cite{Fei97}, called below the $d^9$
model. This model was introduced long ago for a perovskite
KCuF$_3$ \cite{Kug82}, a strongly correlated system with a single
hole within degenerate $e_g$ orbitals at each Cu$^{2+}$ ion. Kugel
and Khomskii showed that orbital order can be stabilized by a
purely electronic superexchange mechanism. This happens for
strongly frustrated orbital superexchange \cite{Ole10}, and
columnar Ising-type of order is obtained \cite{Cin10} in the
two-dimensional quantum compass model. This model exhibits
nontrivial symmetry properties which may be employed to perform
efficient calculations for square compass clusters \cite{Brz10}.

Orbital order occurs in a number of compounds with active orbital
degrees of freedom, where strong Coulomb interaction localizes
electrons (or holes) and gives rise to spin-orbital superexchange
\cite{Kha05}. When spin and orbital pseudospins couple to each
other, their order is usually complementary --- alternating
orbital (AO) order accompanies ferromagnetic (FM) spin order, and
ferro-orbital (FO) order coexists with antiferromagetic (AF) spin
order. However, the above Goodenough-Kanamori rules, see also
\cite{Ole10}, are not satisfied in cases when spin-orbital
entanglement (SOE) dominates \cite{Ole06}, as in the
spin-orbital $d^1$ model on a triangular lattice \cite{Cha11}.

The spin-orbital superexchange KK model for Cu$^{2+}$ ($d^9$) ions
in KCuF$_3$ with $S=1/2$ spins and $e_g$ orbitals described by
$\tau=1/2$ pseudospin was derived from the degenerate Hubbard
Hamiltonian with hopping $t$, intraorbital Coulomb interaction $U$
and Hund's exchange $J_H$ \cite{Ole00}. It describes the
Heisenberg SU(2) spin interactions coupled to the orbital problem
by superexchange $J=4t^2/U$,
\begin{equation} \label{hamik}
{\cal H}=-\frac{1}{2}J\!\!\sum_{\langle ij\rangle||\gamma}
\left\{\left(r_1\,\Pi^t_{\langle ij\rangle}+r_2\,\Pi^s_{\langle
ij\rangle}\right)
\left(\frac{1}{4}-\tau^{\gamma}_i\tau^{\gamma}_j\right)\right.
+ \left. r_3\,\Pi^s_{\langle
ij\rangle}\left(\frac{1}{2}-\tau^{\gamma}_i\right)
\left(\frac{1}{2}-\tau^{\gamma}_j\right)\right\}
-E_z\sum_{i}\tau_i^z\,.
\end{equation}
where $\{r_1,r_2,r_3\}$ depend on $\eta\equiv J_H/U$ \cite{Ole00},
and $\gamma=a,b,c$ is the bond direction. In a bilayer two $ab$
planes are connected by interlayer bonds along the $c$ axis
\cite{Brz11} (a monolayer has only bonds within a single $ab$
plane). Here
\begin{equation}
\label{project} \Pi_{\langle ij\rangle}^{s}=\frac{1}{4}-{\bf
S}_i\cdot{\bf S}_{i+\gamma}, \hskip .7cm \Pi_{\langle
ij\rangle}^{t}=\frac{3}{4}+{\bf S}_i\cdot{\bf S}_{i+\gamma},
\end{equation}
are projection operators on a triplet (singlet) configuration on a
bond $\langle ij\rangle$, and $\tau^{\gamma}_i$ are the orbital
operators for bond direction $\gamma=a,b,c$. They are defined in
terms of Pauli matrices $\{\sigma^x_i,\sigma^z_i\}$ as follows:
\begin{eqnarray}
\tau^{a(b)}_i\equiv\frac{1}{4}\,(-\sigma^z_i\pm\sqrt{3}\sigma^x_i),
\hskip .5cm \tau^c_i=\frac{1}{2}\,\sigma^z_i.
\end{eqnarray}
Finally, $E_z$ is the crystal-field splitting which favors either
$x\equiv x^2-y^2$ (if $E_z>0$) or $z\equiv 3z^2-r^2$ (if $E_z<0$)
orbitals occupied by holes. Thus the model Eq. (1) depends on two
parameters: $E_z/J$ and $\eta$.

The spin-orbital model Eq. (\ref{hamik}) describes also CuO$_2$
planes in La$_2$CuO$_4$, where indeed $U\gg t$ and large
$E_z/J_H\simeq 0.27$ favors holes within $x$ orbitals
\cite{Ole00}. The superexchange between Cu$^{2+}$ ions $\sim
0.127$ eV reproduces there the experimental value. In this paper
we consider the model Eq. (\ref{hamik}) for K$_3$Cu$_2$F$_7$
bilayer and K$_2$CuF$_4$ monolayer compound where nearly 
degenerate $e_g$ orbitals are expected. It has been shown that the
magnetic state of K$_3$Cu$_2$F$_7$ is described by the interlayer 
valence bond (VB) phase \cite{Man07} with AO configuration
whereas K$_2$CuF$_4$ undergoes a pressure driven phase transition 
from the FM phase with alternating orbitals at low pressure to the AF phase
with with $x$ orbitals uniformly occupied.


The simplest approach is a single-site mean field (MF) approximation 
applied to the model Eq. (\ref{hamik}). It excludes any spin 
fluctuations so the spin projectors $\Pi^{t(s)}_{\langle ij\rangle}$ 
($\Pi^s_{\langle ij\rangle}$) can be replaced by their mean values
depending on the assumed magnetic order.
In the orbital sector we apply then the MF decoupling for the
products $\{\tau_i^{\gamma}\tau_{i\pm\gamma}^{\gamma}\}$ along the
axis $\gamma$:
\begin{equation}
\tau_i^{\gamma}\tau_{i\pm\gamma}^{\gamma}
\simeq \langle \tau_i^{\gamma} \rangle \tau_{i\pm\gamma}^{\gamma}+
\tau_i^{\gamma}\langle \tau_{i\pm\gamma}^{\gamma}\rangle -
\langle \tau_i^{\gamma} \rangle\langle \tau_{i\pm\gamma}^{\gamma} \rangle.
\end{equation}
As order parameters we take $t^a\equiv\langle\tau_1^{a}\rangle$
and $t^c\equiv\langle\tau_1^{c}\rangle$ for a chosen site $i=1$
(which is sufficient in orbital sector as $t^b=-t^a-t^c$) and we
assume two orbital sublattices: each neighbor of the site $i$ is
rotated by $\pi/2$ in the $ab$ plane meaning that
$\langle\tau_{i+\gamma}^{a(b)}\rangle=t^{b(a)}$. The
self-consistency equations can be solved analytically 
and the phase diagram can be 
obtained by comparing the ground state energies for different
points in the $(E_z/J,\eta)$ plane (see Ref. \cite{Brz11}). 
One finds two classes of solutions: ($i$) uniform orbital configurations 
($t^c=\pm 1/2$, $t^{a(b)}=\mp 1/4$) for global FO order, and ($ii$) 
nontrivial AO order with orbitals staggering from site to site in 
$ab$ planes.




\begin{figure}[t!]
\begin{center}
    \includegraphics[height=5.8cm]{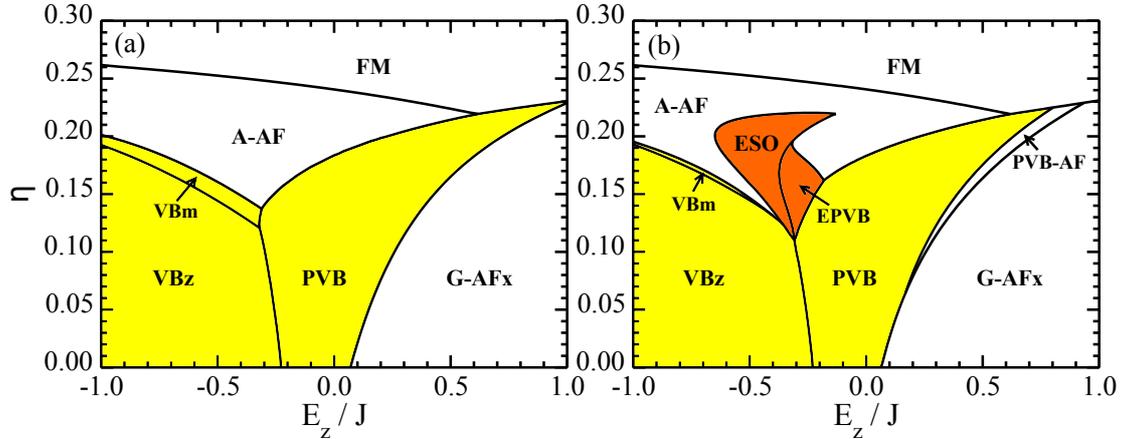}
\caption{
Cluster MF phase diagrams of the $d^9$ bilayer model (1) obtained
with: (a) factorizable SO mean field, and (b) 
independent, potentially non-factorizable SO mean field.
Light shaded (yellow) areas mark singlet phases with spin
disorder and dark (orange) shading indicates phases with SO
entanglement.
}
    \label{diags}
\end{center}
\end{figure}


In a better cluster MF (or Bethe-Peierls-Weiss) approach,
introduced to capture the effects of quantum fluctuations, one
divides the bilayer square lattice into separate cubes containing
8 sites each and treats the bonds inside a cube exactly, and the
bonds connecting different cubes in MF. This approach has at least
three advantages over the single-site MF: ($i$) spins can
fluctuate, ($ii$) elementary cell can double, and ($iii$) we can have 
independent spin-orbital order parameter. The MF leads in a cluster 
to five order parameters: magnetic $ s\equiv \langle S^z_1\rangle$,
orbital $t^{a(b)}\equiv \langle \tau^{a,b}_1\rangle$, and 
spin-orbital $v^{a(b)}\equiv \langle S^z_1\tau^{a,b}_1\rangle$.
The self-consistency equations 
can be solved only numerically by  iterative Lanczos diagonalization 
of a cluster followed by the update of
the mean fields (for details see Ref. \cite{Brz11}). 
To capture the effect of the SOE we first obtain the phase
diagram assuming $v^{a(b)}\equiv s t^{a,b}$ (factorizable 
SO mean field)-- see Fig. \ref{diags}(a) and then do the 
calculation with true $v^{a(b)}$ to get Fig. \ref{diags}(b).


Including spin fluctuations in this approach stabilizes
three valence bond phases absent in the single-site approach.
These are VB$z$, VB$m$ and PVB phases, see Fig. \ref{diags}(a).
The VB$z$ replaces ordinary antiferromagnet in the 
negative $E_z$ region of the phase diagram and 
involve interlayer singlets accompanied by FO$z$ configuration
which can smoothly evolve for growing $\eta $ towards AO 
configuration in the VB$m$ being the phase observed in 
K$_3$Cu$_2$F$_7$ by
Manaka {\it et al.\/} \cite{Man07}. Here we explain it for
realistic $\eta\simeq 0.14$. In the PVB (plaquette VB) phase spin singlets
are pointing uniformly in $a$ or $b$ direction within the cluster and the
clusters form a checkerboard pattern so the unit cell is doubled.
Other phases, i.e. FM, $G$-AF$x$ and $A$-AF, exhibit long
range magnetic order and can be obtained in ordinary MF
approach (see Ref. \cite{Brz11}).


A different class of phases stems from the non-factorizable 
SO mean field and involves SOE --- these are the ESO,
EPVB and PVB-AF phases shown in Fig. \ref{diags}(b) .
The PVB-AF phase connects PVB and $G$-AF
phases by second order phase transitions and is characterized by
fast changes in orbital order and appearance of global
magnetization. The ESO phase has no magnetization and weak FO order. 
When $E_z$ grows, the ESO phase does change continuously into the 
EPVB configuration, being an entangled precursor of the PVB phase
with highly non-uniform AF magnetization. As shown in Ref. \cite{Brz11},
both ESO and EPVB are characterized by strong on-site SOE
defined as $r^{a,b}\equiv v^{a,b}-st^{a,b}$.

\begin{figure}[t!]
\begin{center}
    \includegraphics[height=5.8cm]{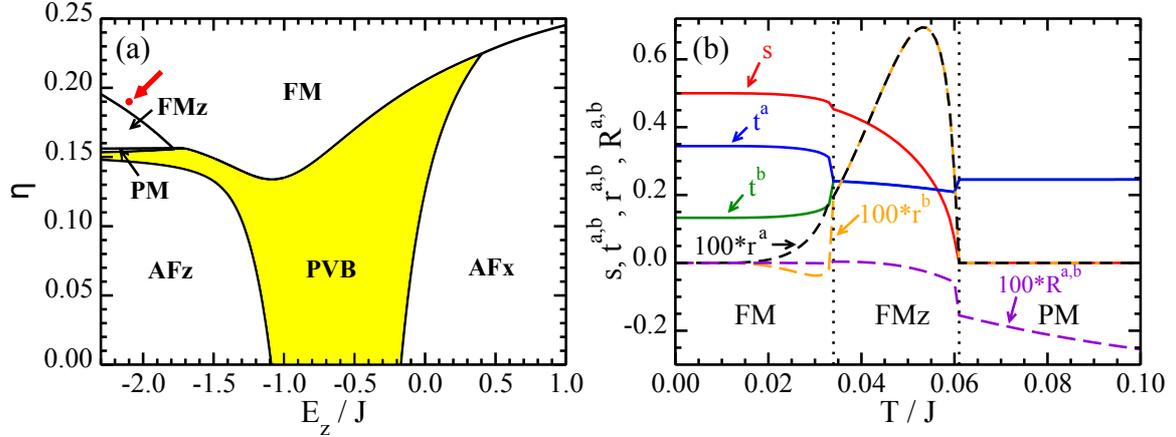}
\caption{Panel (a)-- phase diagram of the $d^9$ monolayer at $T=0$; 
for a more complete phase diagram see \cite{Brz12}. 
Shaded (yellow) area marks singlet phase with spin
disorder. Panel (b)-- spin $s$ and orbital $t^{a,b}$ order parameters,
on-site $r^{a,b}$ and bond $R^{a,b}$ SOE (multiplied by $100$) 
for point $(E_z=-2.1 J, \eta =0.19)$, marked with red arrow in panel (a), 
as functions of temperature. Thermal evolution of the system involves 
intermediate FM$z$ phase before reaching PM phase.}
    \label{diags2}
\end{center}
\vskip -.1cm
\end{figure}

Using the same cluster MF approach as above one can easily study
the properties of the KK model for a single layer at zero and finite
temperature $T$. At $T=0$ one finds the phase diagram of the 
form shown in the Fig. \ref{diags2}(a). In the low-$\eta$ region
it involves only AF and PVB phases and the VB$z$ is replaced
for AF$z$ in the absence of the second layer. The PVB
area is strongly elongated in the horizontal directions
and caps AF$z$ down to $E_z<-2J$ where surprizingly one finds
a narrow stripe of a paramagnetic (PM) phase.
This stripe separates PVB from FM$z$ phase (FM phase 
with FO$z$ order), and shows that frustration in a 
single layer system is bigger than in a bilayer one
where paramagnetizm is absent in $T=0$. 

In the Fig. \ref{diags2}(b) we show the thermal evolution
of spin and orbital order parameters and SOE for a  
point $(E_z=-2.1 J, \eta =0.19)$ lying in the FM area of the
monolayer phase diagram. This evolution involves passage
through the FM$z$ phase and ends in the PM phase.
The FM magnetization $s$ exhibits critical behavior
staying close to $1/2$ in the FM phase and dropping 
quickly to zero at the FM$z$-PM phase transition.
The orbital $t^{a,b}$ show inverse bifurcation 
at the FM-FM$z$ phase transition being the end
of the AO order but remain non-zero in the paramagnet
which is induced by finite $E_z$. The on-site SOE 
represented by $r^{a,b}$ are small everywhere but in 
FM$z$ they become relatively large which lets us 
think of FM$z$ as an entangled phase.
The bond SOE defined by $R^{a,b}\equiv \langle {\bf S}_1{\bf S}_i
\tau^{a,b}_1\tau^{a,b}_i \rangle-\langle {\bf S}_1{\bf S}_i
\rangle\langle\tau^{a,b}_1\tau^{a,b}_i \rangle$ with
$\langle 1i\rangle||a(b)$
take considerable values only in the PM phase
and shows rather purely statistical covariance
than a true quantum effect.

Summarizing, we have shown that spin-orbital entanglement leads to
exotic types of order which are stabilized by {\it quantum
fluctuations} in a bilayer and by {\it thermal fluctuations} in a
monolayer. They emerge from highly frustrated spin-orbital 
superexchange and could be discovered only within a cluster mean
field approach.

\section*{Acknowledgments}

We acknowledge support by the Foundation for Polish Science (FNP) 
and by the Polish Ministry of Science and Higher Education under 
Project No. N202~069639.

\end{document}